\documentstyle[epsfig,procl]{article}

\def\be{\begin{equation}}

\def\ee{\end{equation}}

\def\bea{\begin{eqnarray}}

\def\eea{\end{eqnarray}}

\newcommand{\jpsi}{{J\!/\!\psi}}

\newcommand{\beq}{\begin{equation}}
\newcommand{\eeq}{\end{equation}}
\newcommand{\beqa}{\begin{eqnarray}}
\newcommand{\eeqa}{\end{eqnarray}}

\begin{document}

\begin{flushright}
\mbox{~}\\[-2cm]
DESY 96-144\\
hep-ph/9609236
\end{flushright}
\vspace{1.5cm}

\title{ONIUM PRODUCTION$^\heartsuit$
}
\footnotetext{$^\heartsuit$Talk given at the DIS'96 Workshop, Rome, 
April '96}

\author{ MATTEO CACCIARI }

\address{Deutsches Elektronen-Synchrotron DESY,\\ D-22603 Hamburg, Germany}




\maketitle\abstracts{
The present status of our understanding of onium production is reviewed.
Different models are described and comparisons of theoretical prediction 
with experimental data are given.}

\section*{\vspace{-.8cm}}

In this talk I will review the present status of our understanding of onium
production. Rather than showing many detailed results I will briefly describe
the different production models which are now being considered and tested
against experimental data. When describing charmonium photoproduction I will
restrict myself to the so-called inelastic domain, since the description of
diffractive production would require a talk by itself.

Any model attempting to describe the production of a heavy
quarkonium must deal with two issues which can usually be kept distinct: the
production of the heavy quark-antiquark pair ($Q\bar Q$)
constituting the quarkonium and their binding into a single physical
long-lived particle.

The details of how these two issues are separated and described will of course
depend on the kind of model we consider, but the following general feature can
be seen to apply: the production of the heavy quark-antiquark pair is
described by all models to take place via a short-distance interaction within
perturbative QCD (pQCD). The binding of the two quarks into a bound state is
on the other hand a longer distance process, and the models usually invoke
non-perturbative effects to take place at this point: this part of the
process is usually parametrized via form factors describing the probability of
the $Q\bar Q$ pair to form the bound state. The degree of rigorousness, 
completeness
and complexity of this part of the description varies greatly from model to
model.

Within the last twenty years or so three main models for quarkonia production
have been proposed and used to produce theoretical prediction: the Colour
Evaporation Model  \cite{CEM} (CEM), the Colour Singlet Model  \cite{CSM} (CSM)
and, quite recently, the Factorization Model  \cite{bbl} (FM). They all follow
the general ``guidelines'' outlined above but do however differ in the details
of the hadronization description.

The CEM rests on duality arguments in assuming that {\sl any} $Q\bar Q$ pair
produced with an invariant mass below that of a $D\bar D$ mesons pair (i.e.
below the open charm threshold, using now charm as example of heavy quark)
will eventually hadronize into a quarkonium state. While being physically
sensible, this model does of course have the big drawback of not being able to
predict the production rate of the single quarkonia states. It is therefore
not very suitable for the study of exclusive final states.

The CSM tries to overcome the difficulty of the CEM in predicting rates for
single states by making a very precise request: the $Q\bar Q$ pair must be
produced in the short-distance interaction with the spin, angular momentum and
colour quantum numbers of the quarkonium. A single phenomenological parameter
will then
parametrize its hadronization into the observable particle. This model is of
course much more predictive: the production rate for a colour-singlet $Q\bar
Q$ state with a given spin and angular momentum can be calculated exactly in
pQCD. Moreover, the phenomenological parameter can be measured, for instance,
in electromagnetic decays and used to make absolute prediction about
production rates. Still, also the CSM has its own drawbacks, which have
eventually led to develop a new model. First of all, the simple minded
factorization ``cross section for producing quarkonium equals cross section
for producing colour-singlet $Q\bar Q$ with the proper spin/angular momentum
quantum numbers times a phenomenological parameter'' is known to fail.
Infrared divergences show up in the calculation of the short distance part for
$P$-wave hadronic decays or production \cite{barbieri}: 
a clear signal that this way of
separating short from long distance dynamics is wrong or at least incomplete.
Secondly, CSM predictions for producing $\jpsi$ and $\psi'$ states at large
$p_T$ have been
found to grossly underestimate, by factors of 30 or so, the experimental data
obtained by the CDF collaboration \cite{cdf} in $p\bar p$ collisions at 
the Tevatron (see also  \cite{mlm} and references therein for a review).

The FM (see also ref.  \cite{bfc} for a recent review)
has been proposed to overcome the first of these two problems: it
extends the CSM by allowing $Q\bar Q$ pairs with spin, angular momentum and
colour quantum numbers different from those of the observed quarkonium to
hadronize into the latter. A general expression for a production cross section
within this model then reads
\beq
d\sigma(H + X) = \sum_n d\hat\sigma(Q\bar Q(n) + X)\langle{\cal O}^H[n]\rangle
\label{eq-fm}
\eeq
Here $d\hat\sigma(Q\bar Q(n) + X)$ describes the short distance production of
a $Q\bar Q$ pair in the colour/spin/angular momentum state $n$, and 
$\langle{\cal O}^H[n]\rangle$, formally a vacuum expectation value of a Non
Relativistic QCD matrix element (see  \cite{bbl} for details), describe the
hadronization of the pair into an observable quarkonium state $H$.
 The cross section is no more given by a single product of a short
distance times a long distance part like in the CSM, but rather by a sum of
such terms. Infrared singularities which show up in some of the short distance
coefficients will be absorbed into the long distance part of other terms,
thereby  ensuring a finite overall result.

\begin{figure}
\begin{minipage}{6cm}
\psfig{figure=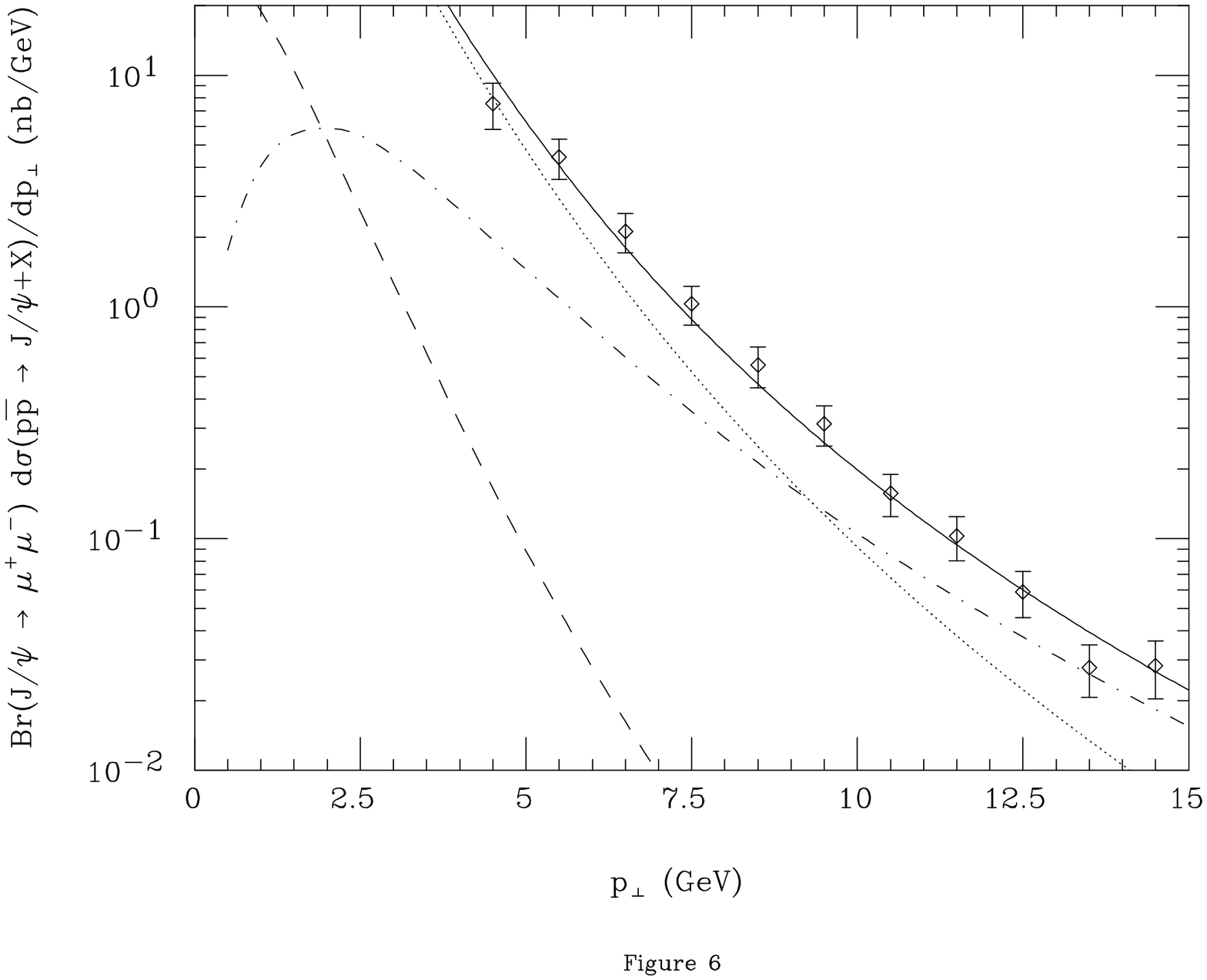,bbllx=70pt,bblly=100pt,bburx=550pt,bbury=770pt,%
   height=6cm,angle=90,clip=}
\caption{$\jpsi$ production at the Tevatron. Dashed line: Color Singlet Model;
dot-dashed line: production via color octet $^3S_1$ states; dotted line:
production via $^1S_0$ and $^3P_J$ states. Non perturbative matrix elements for
octet states fitted to data. Figure from Cho and Leibovich, ref.
\protect \cite{cdfoctet}\label{cdf-fig}}
\end{minipage}
\hspace{.5cm}
\begin{minipage}{5cm}
\psfig{figure=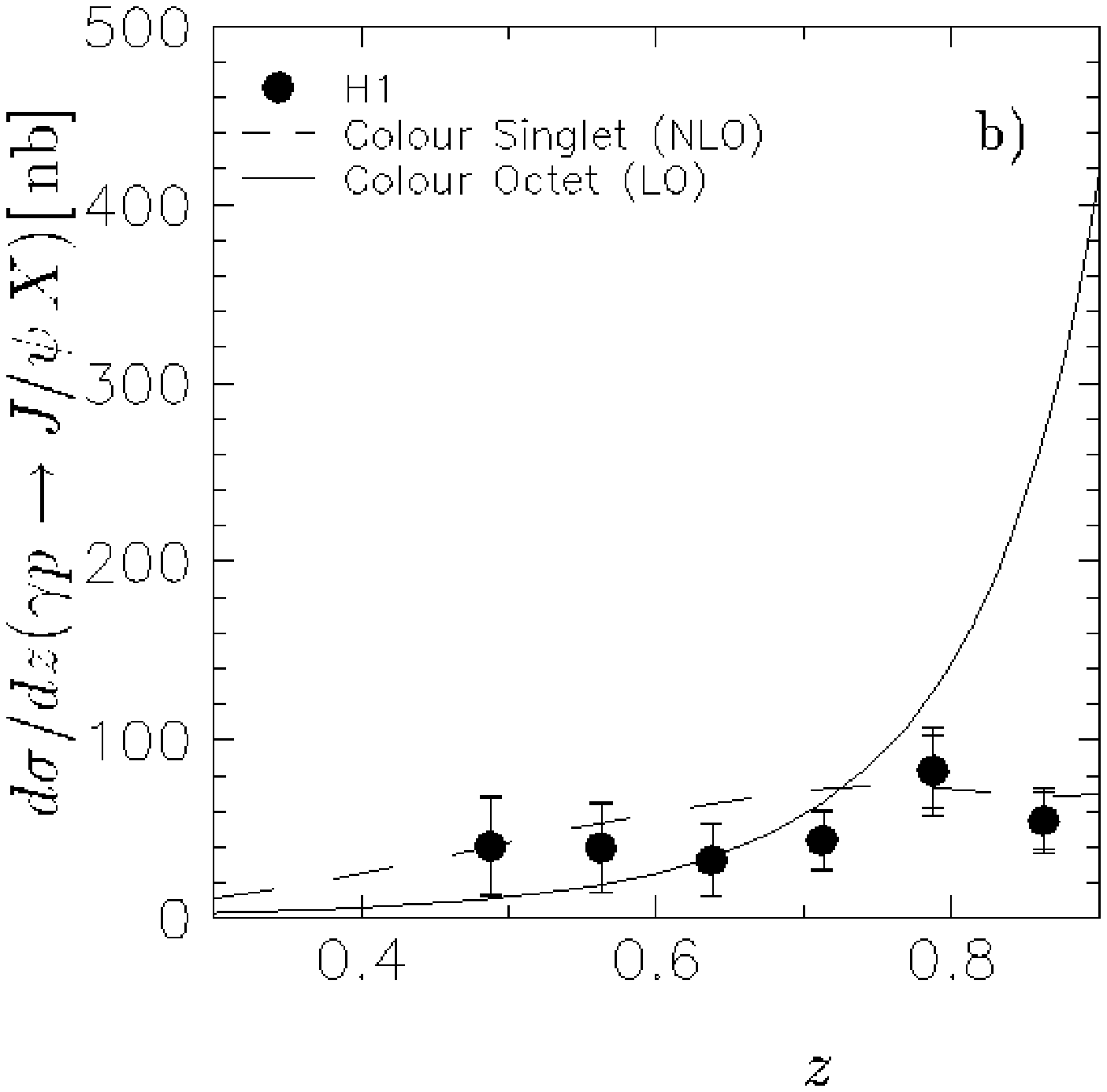,bbllx=90pt,bblly=190pt,bburx=500pt,bbury=600pt,%
   height=5cm,clip=}
\caption{$\jpsi$ production in $\gamma p$ collision at HERA. Dashed line: 
NLO Color Singlet Model; full line: color octet contributions. Plot from
ref.\protect \cite{h1}
\label{h1-fig}}
\end{minipage}
\vspace{-.45cm}
\end{figure}

It is obvious that the FM, by extending the CSM, recovers some of the features
of the CEM: the state of the $Q\bar Q$ pair prepared by the short distance
part of the interaction is no more so strongly restricted. This is in
agreement with the idea that hadronization is a long-distance/long-time scale
process: different  quantum states have the time to evolve into a
physical quarkonium state after their production in a short distance process,
though this evolution is of course suppressed with respect to that of a colour
singlet pair with the appropriate quantum numbers \footnote{Non Relativistic
QCD actually allows one to put these relative suppressions on a more
quantitative ground. See ref. \cite{bbl} for details.}. To suppress them
completely, like in the CSM, is likely to lead to too small cross sections
whenever these states can be copiously produced in the short distance
interaction in comparison with the colour singlet ones. This is the case for
large $p_T$ $\jpsi$ and $\psi'$ production at the Tevatron: it was
found \cite{cdfoctet} that
gluons production and their subsequent fragmentation into a colour octet
$^3S_1$ $c\bar c$ pair (which will eventually hadronize into a physical
quarkonium) can successfully describe the experimental data, previously greatly
underestimated by the CSM (see fig. \ref{cdf-fig}).

The success of the Factorization Model in the description of quarkonium
production at the Tevatron must of course be challenged by comparing its
predictions to experimental data from other reactions, like $e^+e^-$ 
or $\gamma p$ collisions. It turns out that using the non-perturbative matrix
elements fitted to the Tevatron data 
we can produce a parameter free prediction for
inelastic $\jpsi$ photoproduction which can be compared with experimental data
obtained at HERA. The result of this calculation \cite{ck} is shown in fig.
\ref{h1-fig}, where the experimental data from the H1 Collaboration \cite{h1}
are also compared to the NLO CSM prediction \cite{mk}. The plot shows that the
color octet contribution overshoots the data in the large-$z$ region,
while the next-to-leading order CSM prediction seems to describe them
well. Taken at its face value this result would point to a non-universality of
the non-perturbative matrix elements fitted to the Tevatron data and hence to a
failure of the FM. On the other hand, many uncertainties can affect the
theoretical predictions both for the Tevatron and HERA: higher orders and higher
twists could significantly change this picture, and a detailed evaluation of
their relevance is so far not available. More work is therefore needed before 
we can safely handle a successful model for quarkonia production.

\vspace{1mm}
\noindent
{\bf Acknowledgments.} It is a pleasure to thank D. Kosower for inviting me 
to give this talk and the
organizers for the wonderful atmosphere surrounding the whole Conference.

\section*{References}

\newcommand{\zp}[3]{Z.\ Phys.\ {\bf C#1} (19#2) #3}
\newcommand{\pl}[3]{Phys.\ Lett.\ {\bf B#1} (19#2) #3}
\newcommand{\plold}[3]{Phys.\ Lett.\ {\bf #1B} (19#2) #3}
\newcommand{\np}[3]{Nucl.\ Phys.\ {\bf B#1} (19#2) #3}
\newcommand{\prd}[3]{Phys.\ Rev.\ {\bf D#1} (19#2) #3}
\newcommand{\prl}[3]{Phys.\ Rev.\ Lett.\ {\bf #1} (19#2) #3}
\newcommand{\prep}[3]{Phys.\ Rep.\ {\bf C#1} (19#2) #3}
\newcommand{\niam}[3]{Nucl.\ Instr.\ and Meth.\ {\bf #1} (19#2) #3}
\newcommand{\mpl}[3]{Mod.\ Phys.\ Lett.\ {\bf A#1} (19#2) #3}

\end{document}